\documentclass[submission, Phys]{SciPost}
\usepackage{graphicx}
\usepackage{enumerate}
\usepackage{amsmath}
\usepackage{amssymb}
\usepackage[T1]{fontenc}
\usepackage{color}

\newcommand{\e}{\mathrm{e}}

\newcommand{\p}{\partial}

\newcommand{\vid}[1]{\langle #1 \rangle}

\renewcommand{\i}{\mathrm{i}}

\begin{document}

\begin{center}{\Large \textbf{
Classical analog to the Airy wave packet
}}\end{center}

\begin{center}
A. Matulis\textsuperscript{1},
A. Acus\textsuperscript{2*}
\end{center}

\begin{center}
{\bf 1} Semiconductor Physics Institute, Center for Physical Scienes
and Technology, Saul\.{e}tekio 3, LT-10222 Vilnius, Lithuania
\\
{\bf 2} Institute of Theoretical Physics and Astronomy, Vilnius University,
Saul\.{e}tekio 3, LT-10222 Vilnius, Lithuania
\\

*arturas.acus@tfai.vu.lt
\end{center}

\begin{center}
\today
\end{center}


\section*{Abstract}
{\bf
The solution of the Liouville equation for the ensemble of free
particles is presented and the classical analog to the quantum
accelerating Airy wave packet is constructed and discussed.
Considering the motion of various classical packets -- with an
infinite and restricted distribution of velocities of particles --
and also the motion of their fronts, we demonstrate in the
simplest and most definite way why the packet can display a more
sophisticated behavior (even acceleration) as compared with a free
individual particle that moves at a fixed velocity. A comparison
of this classical solution with the quantum one in the Wigner
representation of quantum mechanics which provides the closest
analogy is also presented.
}


\section{Introduction}

Due to dispersion in the Schr\"{o}dinger equation, wave packets usually
spread out during their propagation at a constant velocity in free space.
However, almost half of a century ago it was shown\cite{berry} that there existed a
specific solution of that equation expressed in the Airy function which on the
one hand did not change its form (that is why it is called a coherent state),
and on the other hand, it demonstrated an accelerating motion.
This exotic solutions have recently aroused some interest\cite{vyas} in
possible applications of the description of non-spreading optical beams\cite{sivil}
and the generation and control of plasma in dielectrics\cite{polyn}.

The above mentioned solution is also interesting from a didactic
point of view because of an obvious  contradiction between the
quantum accelerating state and the classical free particle motion
corresponding to it. Usually this contradiction is explained on
the basis of the wave properties of the packet. In our view the
main reason for that is statistical nature of quantum
mechanics. If this is the case such strange packets should exist
in classical mechanics as well, if particles were described
statistically using the Liouville equation. The aim of this paper
is to present solutions of that equation and to discuss the
classical analog to the accelerating quantum coherent state
corresponding to a free particle. To our mind this classical problem
is much simpler as against the quantum one, and consequently it
may be helpful in understanding quantum mechanical problems
better. In addition, we show that the analogy between the
classical and quantum descriptions of the above mentioned
sophisticated packets is quite close if the density matrix
equation in the Wigner representation of quantum mechanics is
used.

The paper is organized as follows. In Sec.~II a description of
the model is provided, convenient dimensionless variables are
introduced and the Liouville equation for the ensemble of free
particles is formulated. A general solution of that equation is
given in Sec.~III. In the following three sections special cases
such as an accelerating packet, a single particle and the packet of a
restricted velocity distribution are discussed. Sec.~VII deals
with the possible accelerating motion of the sharp front. In
Sec.~VIII classical solutions obtained are compared with the
quantum ones given in the Wigner representation of quantum
mechanics. Our conclusions are presented in Sec.~IX.

\section{Model}

The quantum wave packet corresponding to the accelerating quantum particle is described
by the following Schr\"{o}dinger equation:
\begin{equation}\label{sr}
  \i\hbar\frac{\p\Psi}{\p t} = -\frac{\hbar^2}{2m}\frac{\p^2\Psi}{\p x^2}
  - max\Psi = 0,
\end{equation}
where $m$ is the particle mass, $a$ is its acceleration, and $\hbar$ is the Plank
constant. To make expressions more transparent we rescale time $t$, coordinate $x$,
and velocity $v$ as follows:
\begin{equation}\label{scale}
  t \to t\left(\frac{\hbar}{ma^2}\right)^{1/3}, \;
  x \to x\left(\frac{\hbar^2}{m^2a}\right)^{1/3}, \;
  v \to v\left(\frac{\hbar a}{m}\right)^{1/3}.
\end{equation}
After this transformation the initial Eq.~(\ref{sr}) can be presented in
a simpler dimensionless form:
\begin{equation}\label{srb1}
  \i\frac{\p\Psi}{\p t} = -\frac{1}{2}\frac{\p^2\Psi}{\p x^2} - x\Psi.
\end{equation}
The dimensional expressions can be easily restored by applying the same
transformation in the opposite direction.

As mentioned in the Introduction Berry and Balazs have shown\cite{berry}
that the Schr\"{o}dinger equation for the free particle
\begin{equation}\label{srb}
  \i\frac{\p\Psi}{\p t} = -\frac{1}{2}\frac{\p^2\Psi}{\p x^2}
\end{equation}
has the solution
\begin{equation}\label{psiairy}
  \Psi(x,t) = \e^{\i t(x-t^2/3)}\mathrm{Ai}\left[2^{1/3}\left(x - t^2/2\right)\right],
\end{equation}
where the symbol $\mathrm{Ai}$ stands for the Airy function\cite{abr}.
The probability corresponding to this wave function
\begin{equation}\label{tiktank}
  P(x,t) = |\Psi(x,t)|^2 = \mathrm{Ai}^2\left[2^{1/3}\left(x - t^2/2\right)\right]
\end{equation}
is a function of the single argument $(x-t^2/2)$ indicating that it is a coherent
state related to the accelerating particle.

Now we are switching over to our main purpose, and we shall discuss the classical
description of a similar motion. Instead of using Newton's equations for the individual
particle we consider the ensemble of free particles and treat them
statistically. The statistical description of classical particles is based on
the Liouville theorem\cite{gold} which states that the distribution function $f(x,v,t)$
of the particles in the phase space (the density of the particles in $(x,v)$
plane in our case)
behaves like ideal liquid which remains constant when moving along the classical
trajectories. This statement leads to the following Liouville equation
\begin{equation}\label{liuv}
  \frac{\p f}{\p t} + v\frac{\p f}{\p x} = 0
\end{equation}
for the distribution function of free particles.
The solution of this equation enables us to express all properties of the
particles via the integrals with the above distribution function. For
instance, the density of the particles as a function of coordinate and time
is given as follows:
\begin{equation}\label{koncentr}
  n(x,t) = \int_{-\infty}^{\infty}dvf(x,v,t).
\end{equation}
This statistical technique is considered to be an analog to the solution
of Newton's equations for individual particles.

\section{Solution of the Liouville equation}

The simplest way to find the solution of Eq.~(\ref{liuv})
corresponding to the accelerating motion is to look for a static solution
in the accelerating frame $\{y,w\}$ which is related to the initial laboratory
frame $\{x,v\}$ by the following equations:
\begin{subequations}\label{labjud}
\begin{eqnarray}
  x &=& y + t^2/2, \\
  v &=& w + t.
\end{eqnarray}
\end{subequations}
In order to obtain the Liouville equation in the new frame it is necessary
to make the following substitution of the derivatives in Eq.~(\ref{liuv}):
\begin{equation}\label{isvpak}
  \frac{\p f}{\p t} \to \frac{\p f}{\p t} - t\frac{\p f}{\p y} - \frac{\p f}{\p w},
  \quad \frac{\p f}{\p x} \to \frac{\p f}{\p y}
\end{equation}
which leads to the following equation:
\begin{equation}\label{liujud}
  \frac{\p f}{\p t} + w\frac{\p f}{\p y} - \frac{\p f}{\p w} = 0.
\end{equation}
Actually it is the same Liouville equation including the additional force term
that appears due to a non-inertial accelerating frame.

According to Courant\cite{kurant}, the first order partial differential equation
can be solved by means of trajectory method. These trajectories (or characteristic
curves) follow from the set of ordinary differential equations which are
composed of Eq.~(\ref{liujud}) coefficients in the following way:
\begin{equation}\label{trajdif}
  \frac{dt}{1} = \frac{dy}{w} = \frac{dw}{-1}.
\end{equation}
In solving these equations we obtain the trajectories
\begin{subequations}\label{trajsys}
\begin{eqnarray}
  y - wt - t^2/2 &=& y_0, \\
  w + t &=& w_0
\end{eqnarray}
\end{subequations}
defined by two constants $y_0$ and $w_0$.
Actually they are trajectories of individual particles if one
describes them by means of Newton's equations.
According to the above trajectory method, Eq.~(\ref{liujud}) means that
the derivative of the distribution function along the trajectories (\ref{trajsys})
is equal to zero.
Therefore the distribution function is constant along each of them, and
the general solution of the Liouville equation (\ref{liujud}) can be given
as any function of those two constants characterizing the trajectory:
\begin{equation}\label{liusprend}
  f = f(y_0,w_0) \equiv f(y - wt - t^2/2,w + t).
\end{equation}
This expression will be used to obtain and discuss all specific
solutions.

\section{Accelerating solution}
\label{sec_accel}

Let us start with the accelerating solution which is an analog to
the quantum coherent state with the probability given by Eq.~(\ref{tiktank}).

\begin{figure}[t]
\begin{center}
\includegraphics[width=40mm]{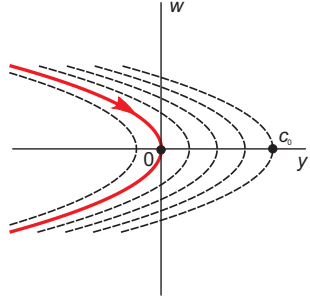}
\end{center}
\caption{\small The set of fixed trajectories in $(y,w)$ plane of
the accelerating frame.}
\label{fig1}
\end{figure}

Eliminating time $t$ from Eqs.~(\ref{trajsys}) we obtain the following equation:
\begin{equation}\label{grpak}
  y + w^2/2 = y_0 + w_0/2 \equiv c_0,
\end{equation}
which defines a set of fixed trajectories in $(y,w)$ plane.
They are shown in Fig.\ref{fig1} by dashed parabolas.
%
The constant $c_0$ indicates the rightmost point of the parabola.
One of them indicated by a red solid curve corresponds to $c_0=0$ value.
Since the solution of the Liouville equation
(\ref{liusprend}) can be chosen as any
function of two constants $y_0$ and $w_0$, we choose it as a function of
their expression given by Eq.~(\ref{grpak}):
\begin{equation}\label{dltatraj}
  f = \delta(y_0 + w_0^2/2) \equiv \delta(y + w^2/2),
\end{equation}
where the symbol $\delta(x)$ stands for the Dirac $\delta$-function.
This function corresponds to the case where all particles
are located only on the red ($c_0=0$) trajectory with constant density.
This distribution can be regarded as a static one only nominally because
from a microscopic point of view all particles move along the trajectories
in the direction shown by the arrow.

\begin{figure}[t]
\begin{center}
\includegraphics[width=40mm]{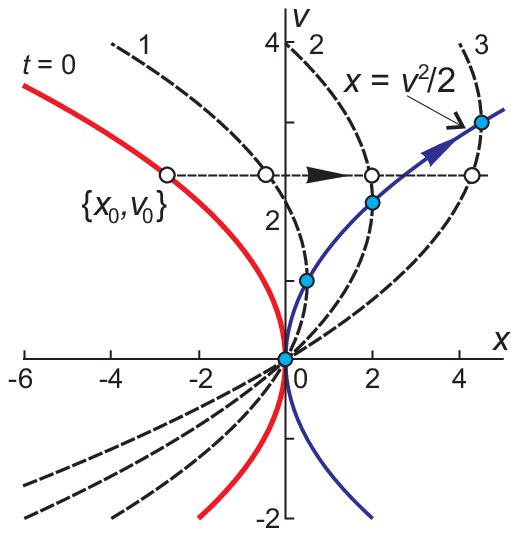}
\end{center}
\caption{\small Motion of the trajectory in the $(x,v)$ plane of the
laboratory frame.}
\label{fig2}
\end{figure}

Now by performing transformation (\ref{labjud})
we obtain the following distribution function in the laboratory frame:
\begin{equation}\label{trajlab}
  f(x,v,t) = \delta\left([x - t^2/2] + [v-t]^2/2\right).
\end{equation}
As time $t$ appeared in the argument of the distribution function, the
trajectories in the laboratory frame are no longer fixed. The motion of the
trajectory corresponding to function (\ref{trajlab}) (to the red trajectory
shown in Fig.~\ref{fig1}) is illustrated in Fig.~\ref{fig2}
%
where time values are indicated by numbers on the curves.
The rightmost points of the moving parabolas are shown by
small solid blue circles. They are located on the solid parabola
\begin{equation}\label{parabol}
  x = v^2/2,
\end{equation}
that is obtained by zeroing both brackets in the argument of function (\ref{trajlab}).
This parabola corresponds to the accelerating particle with
\begin{equation}\label{grpak1}
  x = t^2/2, \quad v = t.
\end{equation}
This accelerating motion reveals itself more clearly
when the density of particles is calculated by means of integral (\ref{koncentr})
with distribution function (\ref{trajlab}), namely,
\begin{equation}\label{konc1}
\begin{split}
  n(x,t) &= \int_{-\infty}^{\infty}dv\delta\left([x - t^2/2] + [v-t]^2/2\right) \\
  &= \frac{1}{|t - v_+|} + \frac{1}{|t - v_-|},
\end{split}
\end{equation}
where zeroes of $\delta$-function argument are as follows:
\begin{equation}\label{nuliai}
  v_{\pm} = t \pm\sqrt{t^2 - 2x}.
\end{equation}
Consequently, the density of particles reads
\begin{equation}\label{konc}
  n(x,t) = \frac{2}{\sqrt{t^2/2 - x}}\Theta(t^2/2 - x),
\end{equation}
where the symbol
\begin{equation}\label{hevis}
  \Theta(x) = \left\{\begin{array}{ll}0, & x < 0; \\ 1 & x\geqslant 0 \end{array}\right.
\end{equation}
stands for the Heaviside step function.

\begin{figure}[t]
\begin{center}
\includegraphics[width=60mm]{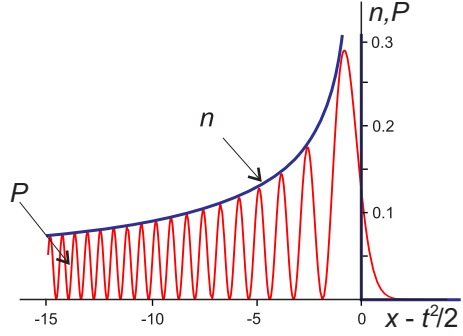}
\end{center}
\caption{\small Accelerating packet: a thin red curve is a quantum mechanical result
according to Eq.~(\ref{tiktank}), a thick blue curve is a classical result according
to Eq.~(\ref{konc}).}
\label{fig3}
\end{figure}

We see that the density demonstrates radical-type singularity that
accelerates in the positive direction of $x$-axis.
It is shown in Fig.\ref{fig3} by a thick solid blue curve as a function of
$(x-t^2/2)$ argument.
%
For comparison the quantum mechanical result (\ref{tiktank}) is indicated by a thin red curve
just below the classical one.
Within the accuracy of normalization
the classical result coincides with the envelope of the asymptote of
the squared Airy function\cite{abr}
\begin{equation}\label{asympt}
  |\mathrm{Ai}(-x)|^2 \sim \frac{1}{\sqrt{\pi x}}\sin^2\left(2x^{3/2}/3+\pi/4\right),
\end{equation}
which confirms that the considered classical problem of free particles is an
analog to the accelerating coherent quantum state.
It leads us to the conclusion that the effect of acceleration is caused
by the statistical properties of the classical system, as well as statistical nature
of the quantum system.

\section{Motion of an individual particle}

Now we shall show that the obtained accelerating state of the ensemble of
particles does not contradict to the motion of a free individual particle
which maintains its constant velocity.

The motion of an individual particle follows from the general solution
of the Liouville equation (\ref{liuv}) when the bellow-presented distribution function
is chosen:
\begin{equation}\label{indiv}
\begin{split}
  f &= \delta(y_0 - x_0)\delta(w_0 - v_0) \\
  &= \delta(y - wt - t^2/2 - x_0)\delta( w + t - v_0).
\end{split}
\end{equation}
It corresponds to the particle which at the moment $t=0$ is located
at the point $\{x_0,v_0\}$.

Using Eqs.~(\ref{labjud}) and going back to the laboratory frame we
obtain the following distribution function:
\begin{equation}\label{indivlab}
  f(x,v,t) = \delta(x - vt - x_0)\delta(v - v_0),
\end{equation}
which shows that the individual particle moves along the trajectory:
\begin{equation}\label{horztraj}
  v = v_0, \quad x = x_0 + v_0t,
\end{equation}
maintaining its initial velocity $v_0$
indicated in Fig.~\ref{fig2} by a thin horizontal dotted line.

Consequently, the effect of acceleration is a pure statistical property of
the ensemble of particles caused by a very specific initial condition
when particles with a constant density are located on the infinite
red trajectory shown in Fig.~\ref{fig1}. This initial distribution
includes particles of infinite velocities, which results in non-integrable
density.

\section{Motion of a packet}

The interesting question is whether it is possible to find an
experimentally realizable initial condition when the
packet demonstrates the accelerating motion.
We shall try to answer this question modifying the distribution function (\ref{dltatraj}),
and adding some localization of velocities close to some average velocity
$v_0$, namely choosing this function as follows:
\begin{equation}\label{trajmod}
\begin{split}
  f(y,w) &= \delta(c_0)\Phi(w_0 - v_0) \\
  &\equiv \delta(y + w^2/2)\Phi(w + t - v_0),
\end{split}
\end{equation}
where the function $\Phi(v)$ stands for some symmetric local distribution
of velocities. For instance, the following Gaussian distribution
\begin{equation}\label{Phidef}
  \Phi(v) = \frac{1}{\sqrt{\pi D}}\e^{-v^2/D}.
\end{equation}
may be chosen.

Going back to the laboratory frame we have
\begin{equation}\label{labmodfk}
  f(x,v,t) = \delta(x+v^2/2-vt)\Phi(v-v_0).
\end{equation}
In a sense we can regard this distribution function as an intermediate
solution between a non physical distribution (\ref{trajlab}) that demonstrates
the accelerating motion and the distribution (\ref{indivlab}) corresponding
to the individual particle that maintains a constant velocity.
This distribution can be generated experimentally by injecting particles at
the point $x=0$ with some retardation depending on the particle velocity.
It might be expected that in changing $D$ value it is possible to change gradually
one type of motion to another one.

Let us check that possibility by calculating the mean coordinate of the packet:
\begin{equation}\label{vidkoord}
\begin{split}
&  \vid{x(t)} 
  = \int_{-\infty}^{\infty}dv\int_{-\infty}^{\infty}dxxf(x,v,t) \\
  &= \int_{-\infty}^{\infty}dv\Phi(v-v_0)
  \int_{-\infty}^{\infty}dxx\delta(x+v^2/2-vt) \\
  &= \int_{-\infty}^{\infty}dv\Phi(v-v_0)\left(vt - v^2/2\right)
  = x_0 + v_0t,
\end{split}
\end{equation}
where
\begin{equation}\label{x0pr}
  x_0 = - v_0^2 - D/4
\end{equation}
denotes the initial mean coordinate of the packet.

We see that the real packet of limited velocities demonstrates
the motion at a fixed average velocity independently of the width $D/4$ of the
packet. This fact has been noticed \cite{pak} before when commenting on the motion
of quantum mechanical packets composed of the superposition of the
above mentioned Airy functions.

\section{The accelerating front}

\begin{figure}[t]
\begin{center}
\includegraphics[width=80mm]{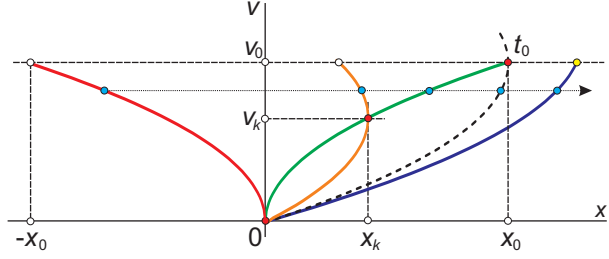}
\end{center}
\caption{\small Motion of the packet with a rectangle distribution
of velocities (\ref{stac}).}
\label{fig4}
\end{figure}

\begin{figure}[t]
\begin{center}
\includegraphics[width=55mm]{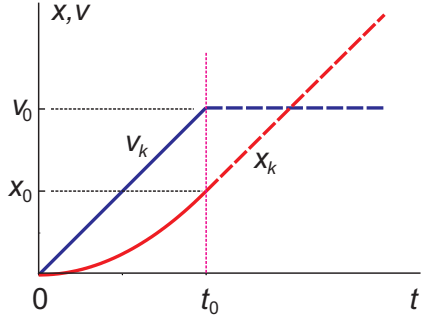}
\end{center}
\caption{\small Motion of the point $\{x_k,v_k\}$ corresponding
to the right front of the packet.}
\label{fig5}
\end{figure}

Although we failed to construct the accelerating packet we show that it is
possible to construct the packet with some accelerating parts, for instance,
with the accelerating front. In order to illustrate this possibility
we consider the packet with a rectangle type distribution of velocities
(the constant velocity in the  region $0 \leqslant w \leqslant v_0$)
on the initial red trajectory shown in Fig.\ref{fig1}:
\begin{equation}\label{stac}
\begin{split}
&  f(y,w,t) = \delta(c_0)\Theta(w_0)\Theta(w_0-v_0) \\
&  = \delta(y + w^2/2)\Theta(w + t)\Theta(v_0 - w - t).
\end{split}
\end{equation}
Using Eqs.~(\ref{labjud}) we go back to the laboratory frame and obtain
the following solution of the Liouville equation:
\begin{equation}\label{reallab}
  f(x,v,t) = \delta(x+ v^2/2a - vt)\Theta(v)\Theta(v_0 - v).
\end{equation}
The evolution of this trajectory is shown in Fig.~\ref{fig4}.
%
Actually it is a horizontal stripe cut out of Fig.~\ref{fig2}, which emerged
owing to the restriction of velocities in the packet.
Every point of the trajectory, say the one indicated by a blue circle,
moves to the right along a thin dotted horizontal line,
and the initial red trajectory moves
to the right and upwards with its rightmost point $\{x_k,v_k\}$
moving along the green parabola $x_k = v_k^2/2$, as it was explained
in Sec.~\ref{sec_accel}. The evolution of the coordinate and
velocity corresponding to this point is shown in Fig.~\ref{fig5}.
%
We see that the acceleration of the front is evident.
But it does not last for ever. At the moment $t_0$ the rightmost point of
the trajectory reaches the limiting velocity $v_0$, and its velocity
no longer increases, as it is seen in Fig.~\ref{fig4}.
The reason for that is trivial: there are no more particles in the
initial distribution with a larger velocity than $v_0$, and the front velocity
stabilizes.

In our opinion this example clearly illustrates the difference between the
motion of an individual particle described by an ordinary differential equation
and the effects of the ensemble of particles described
by a partial derivative equation.
The latter one has many more degrees of freedom,
and therefore it has many more various solutions,
even some exotic ones. In the considered case the acceleration of the front takes place
not only because of acceleration of individual particles but also because of their collective
motion during which faster particles injected later overtake the slower ones,
and consequently, the impression of the accelerating front forms.
There are more effects of that type in physics. It is worth mentioning the
difference in phase and group velocities of the wave in the dispersive media,
or precursors\cite{mat15} moving faster than the wave packets.

\section{Density matrix in the Wigner representation}

The analogy between the classical solution of the Liouville equation and the accelerating
coherent quantum mechanical state that we demonstrated in the previous sections
becomes even closer if the quantum state is described by a density matrix.
It is known that the quantum mechanical description can be based not only on the
wave function $\Psi(x,t)$ which satisfies the Schr\"{o}dinger equation
\begin{equation}\label{sredequ}
  \i\frac{\p}{\p t}\Psi = \hat{H}\Psi,
\end{equation}
but also on the density operator $\hat{\rho}$ that satisfies the quantum Liouville
(or von Neumann) equation
\begin{equation}\label{master}
  \i\frac{\p}{\p t}\hat{\rho} = [\hat{H},\hat{\rho}].
\end{equation}
When comparing Eqs.~(\ref{srb}) and (\ref{sredequ}) we see that in our case of dimensionless
variables the free particle Hamiltonian reads
\begin{equation}\label{ham}
  \hat{H} = -\frac{1}{2}\frac{\p^2}{\p x^2}.
\end{equation}
In the coordinate representation the density operator is usually presented
by the density matrix which in the case of a pure state (in the case where
the quantum system is characterized by the wave function) reads
\begin{equation}\label{tm}
  \hat{\rho} \to \rho(x_1,x_2,t) = \Psi(x_1,t)\Psi^*(x_2,t).
\end{equation}

In 1932 Wigner\cite{Wigner} proposed to replace this density matrix with the function
\begin{equation}\label{wdf}
  F(x,v,t) = \int_{-\infty}^{\infty}d\xi\e^{\i\xi v}\rho(x-\xi/2,x+\xi/2,t),
\end{equation}
which depends on the mean coordinate $x = (x_1+x_2)/2$, and is the
Fourier transform
with respect to the relative coordinate $\xi = x_2 - x_1$.
All other operators of quantum mechanics have to be changed in the same way.
It should be noted that we still use the same dimensionless variables.
That is why there is no Plank constant in the above transformation,
and the velocity $v$ coincides with the dimensionless momentum
$p \to (\hbar m^2a)^{1/3}p$. The equations that are obtained in this way are called
the Wigner representation of quantum mechanics\cite{grot,mat85}.
Its main advantage is that in this representation quantum symbols
and equations are quite similar to their classical counterparts.
For instance, the coordinate and momentum operators convert themselves just
to the numbers similar to the coordinate $x$ and the momentum $v$ used in classical
mechanics. Their mean values are expressed in terms of simple integrals with
the Wigner distribution function $F(x,v,t)$:
\begin{subequations}\label{klxpvid}
\begin{eqnarray}
  \vid{x} &=& \frac{1}{2\pi}\int_{-\infty}^{\infty}dx\int_{-\infty}^{\infty}dvxF(x,v,t), \\
  \vid{v} &=& \frac{1}{2\pi}\int_{-\infty}^{\infty}dx\int_{-\infty}^{\infty}dvvF(x,v,t).
\end{eqnarray}
\end{subequations}
The density of the particles is also given by the integral similar to its
classical counterpart
\begin{equation}\label{qkonc}
  N(x,t) = \frac{1}{2\pi}\int_{-\infty}^{\infty}dvF(x,v,t).
\end{equation}
Some price has to be paid for this simplicity and clearness:
the product of two operators and their commutator have more
complicated expressions.
Hence, if in the Wigner representation the quantum mechanical operators $\hat{A}$ and
$\hat{B}$ are replaced
with the functions $a(x,v)$ and $b(x,v)$, correspondingly,
their product and commutator have to be presented as follows:
\begin{subequations}\label{opersand}
\begin{eqnarray}
\label{opersand1}
  \hat{A}\cdot\hat{B} &\leftrightarrows &
  \exp\left[\frac{1}{2\i}\left(\frac{\p^{(a)}}{\p v}\frac{\p^{(b)}}{\p x}
  -\frac{\p^{(a)}}{\p x}\frac{\p^{(b)}}{\p v}\right)\right]ab, \phantom{mmm} \\
\label{opersand2}
  \left[\hat{A},\hat{B}\right] &\leftrightarrows &
  \frac{2}{\i}\sin\left[\frac{1}{2}\left(\frac{\p^{(a)}}{\p v}\frac{\p^{(b)}}{\p x}
  -\frac{\p^{(a)}}{\p x}\frac{\p^{(b)}}{\p v}\right)\right]ab. \phantom{m}
\end{eqnarray}
\end{subequations}
These equations are merely mnemonic rules presenting the Taylor expansions of
the derivatives where indexes show which function ($a$ or $b$) they have to
be applied to.

Although the sinus and exponent Taylor expansions are infinite, very often
they break off when applied to polynomial-type functions.
Such is the free particle Hamiltonian (\ref{ham})
which becomes a quadratic function of the momentum $H= v^2/2$ in the Wigner representation.
Due to this the sinus function in Eq.~(\ref{opersand2}) can be replaced with
its argument, and the commutator can be rewritten as follows:
\begin{equation}\label{wtmlg}
\begin{split}
  [H,\rho] &\leftrightarrows
  \frac{2}{\i}\sin\left[\frac{1}{2}\left(\frac{\p^{(v)}}{\p v}\frac{\p^{(F)}}{\p x}
  -\frac{\p^{(v)}}{\p x}\frac{\p^{(F)}}{\p v}\right)\right]\frac{v^2}{2}F \\
  &= -\i\left(\frac{\p^{(v)}}{\p v}\frac{\p^{(F)}}{\p x}
  -\frac{\p^{(v)}}{\p x}\frac{\p^{(F)}}{\p v}\right)v^2F \\
  &= -\i\frac{\p^{(v)}v^2}{\p v}\frac{\p^{(F)}F}{\p x}
  = -\i v\frac{\p F}{\p x}.
\end{split}
\end{equation}
Inserting this expression into Eq.~(\ref{master}) we finally obtain the
following quantum Liouville equation in the Wigner representation:
\begin{equation}\label{ldwflg}
  \frac{\p F}{\p t} + v\frac{\p F}{\p x} = 0,
\end{equation}
which coincides exactly with the classical Eq.~(\ref{liuv}).

Consequently, in the case of simple systems (that of free particles)
the density matrix in the Wigner representation satisfies the classical
Liouville equation, and the quantum effects may reveal themselves
only due to additional restrictions such as the boundary or initial conditions
to be satisfied by the density matrix.

In order to understand better the difference between the classical and quantum
solution of the Liouville equation we compute the Wigner function (\ref{wdf})
in our case of the coherent state described by the wave function of Airy
type (\ref{psiairy}):
\begin{equation}\label{vigatv}
\begin{split}
  F(x,v,t) = \int_{-\infty}^{\infty}d\xi & \e^{\i\xi(v-t)}
  \mathrm{Ai}\left[2^{1/3}\left(X - \xi/2\right)\right] \\
  &\times\mathrm{Ai}\left[2^{1/3}\left(X + \xi/2\right)\right],
\end{split}
\end{equation}
where
\begin{equation}\label{Xdef}
  X = x - t^2/2.
\end{equation}
This integral can be calculated analytically taking advantage of the integral
representation of the Airy function:
\begin{equation}\label{airy}
  \mathrm{Ai}(x) = \frac{1}{2\pi}\int_{-\infty+\i\alpha}^{\infty+\i\alpha}
  du \e^{\i(xu + u^3/3)},
\end{equation}
where the contour of integration is shifted up
in the complex $u$ plane by small quantity $\alpha$ to insure convergence
of the integral.
Now denoting $P = v - t$, $a = 2^{1/3}$, and omitting the integral limits
(they are the same as in integral (\ref{airy})) we write down the Wigner
function as follows:
\begin{equation}\label{Vgfk}
\begin{split}
&  F = \frac{1}{(2\pi)^2}\int_{-\infty}^{\infty}d\xi\e^{\i(P - au/2 + as/2)\xi} \\
&\phantom{mmm} \times \int du\e^{\i u^3/3 + \i aXu}\int ds\e^{\i s^3/3 + \i aXs} \\
  &= \frac{1}{2\pi}\int du\e^{\i u^3/3 + \i aXu}
  \int ds\e^{\i s^3/3 + \i aXs} \\
&\phantom{mmm} \times   \delta(P - au/2 + as/2) \\
  &= \frac{1}{\pi a}\int du\e^{\i u^3/3 + \i aXu}
  \e^{\i (u-P/a)^3/3 + \i aX(u-P/a)}.
\end{split}
\end{equation}
Finally substituting $u = (\eta + P)/a$ we rewrite this integral as
\begin{equation}\label{integ}
  F = \frac{a}{2\pi}\int d\eta\e^{\i[\eta^3/3 + 2\eta(P^2/2+X)]},
\end{equation}
which according to Eq.~(\ref{airy}) enables this Wigner function
to be expressed in terms of the Airy function:
\begin{equation}\label{coherfin}
  F(x,v,t) =  2^{1/3}\mathrm{Ai}\left(2[(v-t)^2/2 + (x-t^2/2)]\right).
\end{equation}
We see that the argument of this function coincides with the argument of
the classical distribution function (\ref{trajlab}) which proves
that the Wigner function of the considered pure coherent Airy state can be
constructed in the same way as the classical solution
of the Liouville equation, namely, as some constant distribution along
the classical trajectories. The only difference is that in the quantum
case the constants characterizing the different trajectories cannot
be chosen arbitrary which indicates a well-known fact that the quantum
solution is stronger correlated as compared with the classical one.
This correlation appears as oscillations, the characteristic period
of which, according to Eq.~(\ref{scale}), is proportional to $\hbar^{2/3}$.
Incidentally, this result illustrates a plain fact
that the limiting transition
of quantum mechanics to classical one at $\hbar\to 0$ is
quite complicated: the quantum oscillations disappear due to an increase in their
frequency rather than due to a decrease in their amplitude.

Meanwhile the density calculated by the integral (\ref{qkonc}) with the Wigner
function (\ref{coherfin}) coincides with the one obtained by Eq.~(\ref{tiktank})
that we have checked numerically.

\section{Conclusion}

We presented a general solution of the Liouville equation describing an
ensemble of free particles. In considering various specific cases we showed that
although individual particles moved at a fixed velocity, the packet
constructed of them can demonstrate quite a different behavior.
If the packet includes the particles with unrestricted velocities it can
even accelerate, and thus, it can become an analog to the
quantum coherent accelerating state described by the Airy function.

The mean coordinate of the packet which is composed of the particles
with restricted velocities do not show any acceleration.
However, even in this case it is possible either to choose the initial condition
or inject particles in a correlated way so that some parts of the packet,
say, its front, should accelerate during the finite time interval.
The essence of this motion is that the particles which are injected later
overtake the slowest ones, and in this way create the image of the
accelerating front.

These conclusions are applied in the case of the quantum free particle
because the density matrix equation in the Wigner representation
for this simple system coincides with the classical Liouville equation.

\section*{Acknowledgements}
This research was partly (A.~Acus) funded by the European Social
Fund under Grant No. 09.3.3-LMT-K-712-01-0051.
The authors are sincerely grateful to Professor K\k{e}stutis Pyragas for a
detailed discussion of the work and his helpful comments.

\bibliography{S1}

\begin{thebibliography}{10}
\providecommand{\url}[1]{\texttt{#1}}
\providecommand{\urlprefix}{URL }
\expandafter\ifx\csname urlstyle\endcsname\relax
  \providecommand{\doi}[1]{doi:\discretionary{}{}{}#1}\else
  \providecommand{\doi}{doi:\discretionary{}{}{}\begingroup
  \urlstyle{rm}\Url}\fi
\providecommand{\eprint}[2][]{\url{#2}}

\bibitem{berry}
M.~V. Berry and N.~L. Balazs,
\newblock \emph{Nonspreading wave packets},
\newblock Am.\ J.\ Phys. \textbf{47}, 264 (1979),
\newblock \doi{10.1119/1.11855}.

\bibitem{vyas}
V.~M. Vyas,
\newblock \emph{Airy wave packets are {P}erelomov coherent states},
\newblock Am.\ J.\ Phys. \textbf{86}, 750 (2018),
\newblock \doi{10.1119/1.5051181}.

\bibitem{sivil}
G.~A. Siviloglou and D.~N. Christodoulides,
\newblock \emph{Accelerating finite energy airy beams},
\newblock Opt. Lett. \textbf{32}, 979 (2007),
\newblock \doi{10.1364/OL.32.000979}.

\bibitem{polyn}
J.~V. M. G. A.~S. P.~Polynkin, M.~Kolesnik and D.~N. Christodoulides,
\newblock \emph{Curved plasma chanel generation using ultraintense airy beams},
\newblock Science \textbf{324}, 229 (2009),
\newblock \doi{10.1126/science.1169544}.

\bibitem{abr}
M.~Abramowitz and I.~A. Stegun,
\newblock \emph{Handbook of mathematical functions},
\newblock National bureau of standards, Washington,
\newblock Sec.~10.4. (1972).

\bibitem{gold}
C.~P. H.~Goldstein and J.~Safko,
\newblock \emph{Classical mechanics},
\newblock Addison Wesley, New York,
\newblock P.~419. (2000).

\bibitem{kurant}
R.~Courant and D.~Hilbert,
\newblock \emph{Methods of Mathematical Physics, volume II, Partial
  differential equations},
\newblock John Wiley \& Sons, New York,
\newblock P.~28. (1989 \S).

\bibitem{pak}
J.~Lekner,
\newblock \emph{Airy wavepacket solutions of the schr\"{o}dinger equation},
\newblock Eur.~J.~Phys. \textbf{30}, L43 (2009),
\newblock \doi{10.1088/0143-0807/30/3/L04}.

\bibitem{mat15}
M.~Z. A.~Matulis and F.~M. Peeters,
\newblock \emph{Wave fronts and packets in 1d models of different
  meta-materials: Graphene, left-handed media and transmission line},
\newblock Phys,~Status Solidi \textbf{10}, 2330 (2015),
\newblock \doi{10.1002/pssb.201552023}.

\bibitem{Wigner}
E.~Wigner,
\newblock \emph{On the quantum correction for thermodynamic equilibrium},
\newblock Phys.~Rev. \textbf{40}, 749 (1932),
\newblock \doi{10.1103/PhysRev.40.749}.

\bibitem{grot}
S.~R. Groot and L.~G. Suttorp,
\newblock \emph{Foundations of Electrodynamics},
\newblock North-Holland publishing company, Amsterdam,
\newblock Ch.~6, p.~317. (1971).

\bibitem{mat85}
A.~Matulis,
\newblock \emph{Quantum nonlinenar oscillator in the wigner representation},
\newblock Lith.~J.~Phys. \textbf{25}, 53 (1986),
\newblock (in Russian).

\end{thebibliography}

\end{document}